\documentclass[aps,reprint]{revtex4-2}
\usepackage{amsmath}
\usepackage{adjustbox}
\usepackage{tikz}
\usepackage{float}
\usepackage{tabularx}
\usepackage{soul}
\usepackage[normalem]{ulem}
\usepackage{url}
\usepackage{graphicx}
\usepackage{xr-hyper}
\usepackage{hyperref}
\usepackage{isotope}
\usepackage{mhchem}
\usepackage{xcolor}
\usepackage{tikz}
\usepackage{float}
\usepackage{tabularx}
\usepackage{makecell}
\usepackage{amsmath}
\usepackage{amssymb}
\usepackage{sidecap}

\usepackage{xcolor}
\definecolor{pastelgray}{rgb}{0.81, 0.81, 0.77}
\definecolor{beaublue}{rgb}{0.9, 0.9, 0.93}

\begin{document}
%\preprint{APS/123-QED}
\title{Source Function from Two-Particle Correlation Through Deblurring}

\author{Pierre Nzabahimana}
 \affiliation{Facility of Rare Isotope Beams and Department of Physics and Astronomy, \\
 	Michigan State University, East Lansing, Michigan 48824, USA}
\author{Pawel Danielewicz}
 \affiliation{Facility of Rare Isotope Beams and Department of Physics and Astronomy, \\
 	Michigan State University, East Lansing, Michigan 48824, USA}

\begin{abstract}

In heavy-ion collisions, low relative-velocity two-particle correlations have been a tool for assessing space-time characteristics of particle emission.  Those characteristics may be cast in the form of a relative emission source related to the correlation function through the Koonin-Pratt (KP) convolution formula that involves the relative wave-function for the particles in its kernel.  In the literature, the source has been most commonly sought by parametrizing it in a Gaussian form and fitting to the correlation function. At times the source was more broadly imaged from the function, still employing a fitting.  Here, we propose the use of the Richardson-Lucy (RL) optical deblurring algorithm for deducing the source from a correlation function.  The RL algorithm originally follows from probabilistic Bayesian considerations and relies on the intensity distributions for the optical object and its image, as well as the convolution kernel, being positive definite, which is the case for the corresponding quantities of interest within the KP formula.

\end{abstract}
\maketitle
%\section{INTRODUCTION}
Correlations of particles emitted from heavy-ion collisions are a powerful tool for learning both about the emitting heavy-ion system and about the subsystem of the two measured particles.  As to the subsystem, it may be possible to learn about the resonances formed and, more generally, about the interaction between the two particles \cite{lednicky_final_1981, morita_lambda_2015, adamczyk_measurement_2015, alice_collaboration_investigation_2023}, that is especially important when one or both of the particles in the subsystem are unstable.  As to the overall heavy-ion system, it may be possible to learn, from correlations, about any developed collective motion and local temperature at emission \cite{danielewicz_collective_1988, pochodzalla_nuclear_1985}.  For low relative velocities within the pair, it may be possible to learn the space-time geometry behind particle emission and, more specifically, the distribution of emission points within the reference system co-moving with the particle pair~ \cite{goldhaber_influence_1960, koonin1977proton, pratt1984pion, bauer1993particle, heinz1996lifetimes, bertsch_meson_1996, BROWN1997252, lisa_femtoscopy_2005}.

The basis for learning about the distribution of emission points from low relative velocity correlations is the so-called Koonin-Pratt (KP) formula that represents the measured correlation function as a convolution of the measured correlation function with a relative distribution of emission points \cite{koonin1977proton, pratt1984pion, BROWN1997252}.  The convolution kernel involves the square modulus of the pair relative wave function with the outgoing boundary condition of the relative momentum determined at detectors.  The formula can be derived through a reduction of the two-particle yield from a collision \cite{danielewicz_formulation_1992}.  Provided there are structures in the square of the relative wave function, changing with the relative momentum, their interplay with the source function can give rise to structures in the measured correlation function, on which source inference relies upon. 

In the literature, the source functions have been most often parametrized, usually in a Gaussian form, and fitted to the correlation data \cite{goldhaber_influence_1960, PhysRevC.33.549, pratt1984pion, boal90, verde_imaging_2002}.  In the measurements, the correlation functions have been usually averaged, at least at some level, over orientations of the relative momentum, and correspondingly the inferred source functions were to represent emission points averaged over orientations of the relative position vector.  However, at times the correlation functions have been measured in a differential manner over angles and three-dimensional Gaussian source shapes have been fitted \cite{lisa_femtoscopy_2005}.  Moreover, the source determination from correlation problem has been recognized as one of the imaging \cite{BROWN1997252} and imaging process of the source function was undertaken without prejudice on the source shape \cite{e895_collaboration_comparison_2003, brown_imaging_2005}.

Here, we return to the problem of source imaging from the correlation measurements, that principally, like elsewhere for imaging, invokes inversion and thus may suffer from instabilities.  Rather than applying an inversion directly, we take inspiration from optical deblurring that is an imaging problem too.  One successful strategy there, that has been already ported into nuclear physics to cope with detector inefficiencies and reaction-plane uncertainties, is the Richardson-Lucy (RL) method \cite{richardson_bayesian-based_1972, lucy_iterative_1974, danielewicz2022deblurring} that relies on the Bayes theorem.  The RL method largely owes its success to the fact that it operates with strictly positive definite quantities, the probabilities.  Conveniently, the corresponding quantities of interest within the KP formula are positive definite, even though the overall meaning of the KP formula differs from that providing context for the RL method.

Experimentally the correlation function $C(\mathbf{q})$ between particles 1 and 2 is defined with
\begin{equation}
C(\mathbf{q}) = R(\mathbf{q}) +1 = \frac{\frac{dN_{12}^6(\mathbf{p}_1,\mathbf{p}_2)}{d^3\mathbf{p_2} \, d^3\mathbf{p_2}}}{\frac{dN_{1}^3(\mathbf{p}_1)}{d^3\mathbf{p}_1}\frac{dN_{2}^3(\mathbf{p}_2)}{d^3\mathbf{p}_2}} \,
\label{Eq630}
\end{equation}
where $\mathbf{q}$ is the relative momentum in the center of mass of the particles with momenta $\mathbf{p}_1$ and  $\mathbf{p}_2$ and the numerator on the r.h.s.\ is the coincidence yield per collision event and the numerator is the product of single-particle yields.  The naive expectation in a heavy-ion collision with a multiparticle final states is that emission is uncorrelated for moderate $q$, i.e., $C \approx 1$ there.  With this, the actual correlation information of interest is a deviation from 1, with the latter isolated as $R$ in the center part of Eq.~\eqref{Eq630}.  In the literature, $R$ is also often referred to as correlation.

On the theoretical side, at $q \longrightarrow 0$, the correlation function may be represented in terms of the KP formula:
\begin{equation}
    C(\mathbf{q}) = \int d^3 r \, |\Psi_\mathbf{q}^{(-)}(\mathbf{r})|^2 \, S(\mathbf{r})
    \equiv \int d^3r \, K(\mathbf{q},\mathbf{r}) \, S(\mathbf{r})
    \, .
    \label{eq:CPsiS}
\end{equation}
Here, $\Psi_\mathbf{q}^{(-)}$ is a 2-particle 
scattering wave function specified with incoming wave boundary conditions and asymptotically representing the center-of-mass relative momentum~$\mathbf{q}$.  Possible spin indices are suppressed at this stage.  The wave function normalization is such that the kernel in the KP relation, $K \equiv |\Psi|^2$, averages to 1 in the asymptotic zone of large~$r$.  The function $S(\mathbf{r})$ is the probability distribution of particles 1 and 2 in their separation $\mathbf{r}$ in their center of mass, for the instant when they separate from the rest of the system and leave for the detectors.  That distribution is normalized to 1, $\int d^3r \, S(\mathbf{r}) = 1$.  With this, for larger $q$, the correlation function $C$ on the l.h.s.\ of Eq.~\eqref{eq:CPsiS} is expected to approach 1.  In fact, the experimental correlation functions~$C$, Eq.~\eqref{Eq630}, are often normalized to 1 at intermediate~$q$, in the context of source $S$ inferences.  When $q$ reaches typical kinematic size of relative momenta in a collision, effects not captured in~\eqref{eq:CPsiS} begin to play a role in the measured correlations and, in particular, effects of reaction plane and momentum conservation.  The ability to learn from the low relative-velocity correlations is further emphasized by subtracting unity from both sides of Eq.~\eqref{eq:CPsiS} and arriving at an equation for the $R$ correlation function \cite{BROWN1997252}:
\begin{equation}
    R(\mathbf{q}) = \int d^3 r \, \Big( |\Psi_\mathbf{q}^{(-)}(\mathbf{r})|^2 - 1 \Big) \, S(\mathbf{r}) \, .
    \label{eq:RKS}
\end{equation}
Provided the interaction within the particle pair is constrained, so that $|\Psi|$ can be faithfully assessed for $q$ and $r$ of interest, then $S$ may be inferred from any structures in $R$.  If the interaction is unknown, but generic assumptions on $S$ can be made, then the KP relation may be used to constrain the interaction between the particles.

The correlation function averaged over directions of $\mathbf{q}$ is related to the source function averaged over the directions of the relative separation $\mathbf{r}$:
\begin{equation}
    C(q) = 4\pi \int_0^\infty dr \, r^2 \, K(q,r) \, S(r) \, ,
    \label{eq:CPsiS1D}
\end{equation}
and
\begin{equation}
    K(q,r) =    \overline{|\Psi^{(-)}_\mathbf{q}(\mathbf{r})|^2}  \, ,
    \label{eq:KqrPsi}
\end{equation}
where the r.h.s.\ is the squared wave function is averaged over orientation of $\mathbf{r}$ relative to $\mathbf{q}$.

As may be apparent in Eqs.~\eqref{eq:CPsiS} and \eqref{eq:RKS}, the inference of $S$ from $C$ represents an imaging problem.  In fact, for neutral pion pairs, with weak strong-interaction effects within the pair ignored, the kernel in \eqref{eq:CPsiS} becomes $K(\mathbf{q}, \mathbf{r}) = 1 + \cos{2 \mathbf{q} \cdot \mathbf{r}}$, so that the correlation $R$ in~\eqref{eq:RKS} becomes a Fourier transform of the source $S$ \cite{BROWN1997252}.  For the general task of imaging, in this work we reach for the Bayesian RL method that was originally developed for deblurring optical images \cite{richardson_bayesian-based_1972, lucy_iterative_1974, vankawala_survey_2015}, but has been by now invoked for nuclear problems bearing similarity to the optical deblurring \cite{vargas_unfolding_2013, danielewicz2022deblurring, Pierre}.  Here, we will carry that method to the application even farther from the method's origins.

In the optical blurring problem, a photon is measured with a property $t'$, while its true property is $t$.  The forward blurring relation, between the distribution 
$\mathcal{F}$ in the true property $t$ and the measured distribution $f$ in the attributed property $t'$, is
\begin{equation}
f(t^{\prime})=\int dt \, P(t^{\prime}|t) \, \mathcal{F}(t) \,  .
\label{Eq1}
\end{equation}
Here, $P(t^{\prime}|t)$ is the conditional probability that a photon with true property $t$ is measured with property $t'$.  When the properties are discretized, such as in attributing the photon to a particular pixel, the relation becomes one in the matrix form between the distribution vectors:
\begin{equation}
f_i=\sum_i P_{ij} \, \mathcal{F}_j \,  .
\label{Eq1}
\end{equation}

A deblurring method, such as RL, seeks to determine the distribution $\mathcal{F}$, when knowing $f$ and $P$.  To arrive at the RL strategy, a backward relation between $f$ and $\mathcal{F}$ is invoked, that involves a conditional probability $Q$ that is complementary to $P$.  Requiring the fulfillment of a Bayesian relation involving $P$ and $Q$, $\mathcal{F}$ is searched through iterations \cite{danielewicz2022deblurring}
\begin{equation}
    \mathcal{F}^{(\mathfrak{r}+1)}_j = \mathcal{F}^{(\mathfrak{r})}_j \,  \frac{\sum_i W_i \frac{f_i}{f_i^{(\mathfrak{r})}} \, P_{ji} }{\sum_i W_i P_{ji}} \equiv \mathcal{F}^{(\mathfrak{r})}_j \, A^{(\mathfrak{r})}_j \, .
    \label{eq:RL}
\end{equation}
Here, $\mathfrak{r}$ is the iteration index, $A^{(\mathfrak{r})}$ is an amplification factor, and $f^{(\mathfrak{r})}$ is prediction for the observation at $\mathfrak{r}$'th iteration:
\begin{equation}
    f^{(\mathfrak{r})}_i=\sum_i P_{ij} \, \mathcal{F}^{(\mathfrak{r})}_j \,  .
\end{equation}
Finally, $W$ is a weight specifying relative importance of the particular data in inferring $\mathcal{F}$.  Errors in the inference of  $\mathcal{F}$ may be assessed by resampling $f$ for the restoration, within the errors of the measurement \cite{Pierre}.

%In the case of an excessive binning in $\mathcal{F}$, as compared to $f$ and/or significant errors in $f$, instabilities can develop in the limit of prolonged iterations for $\mathcal{F}$.  These can be tamed by introducing a stabilizing factor $I$:

If we compare Eqs.~\eqref{eq:CPsiS} or \eqref{eq:CPsiS1D} to \eqref{Eq1}, we can see a connection in the analogous mathematical structure.  Moreover, each of the quantities in \eqref{eq:CPsiS} has a probabilistic interpretation, though only $S$ ties directly to $\mathcal{F}$ in  \eqref{Eq1}.  We will primarily rely on the analogous mathematical structure in \eqref{eq:CPsiS} or \eqref{eq:CPsiS1D} and \eqref{Eq1} and attempt to use the RL method to deduce $S$.  The weights $W$ in \eqref{eq:RL} can serve to focus attention on the region of relative momenta in the correlation function dominated by the interplay of the particles with each other.  

As illustration in this paper, we choose correlations between deuteron and alpha particles.  For this particle combination, scattering phase shifts have been measured \cite{mcintyre_phase_1967} and phenomenological potentials were developed allowing for calculations of scattering wave functions \cite{mcintyre_phase_1967, PhysRevC.33.549}.  Also correlation functions between those particles have been measured \cite{PhysRevC.36.2297, chitwood_final-state_1985, GHETTI2006307}.  Besides feasibility of the source inference with a deblurring algorithm, we will consider practicalities of the inference, such as the binning decisions for $C$ and $S$, errors of inference and impact of detector resolution.

\begin{figure}
    \centering
    \includegraphics[scale=.43]{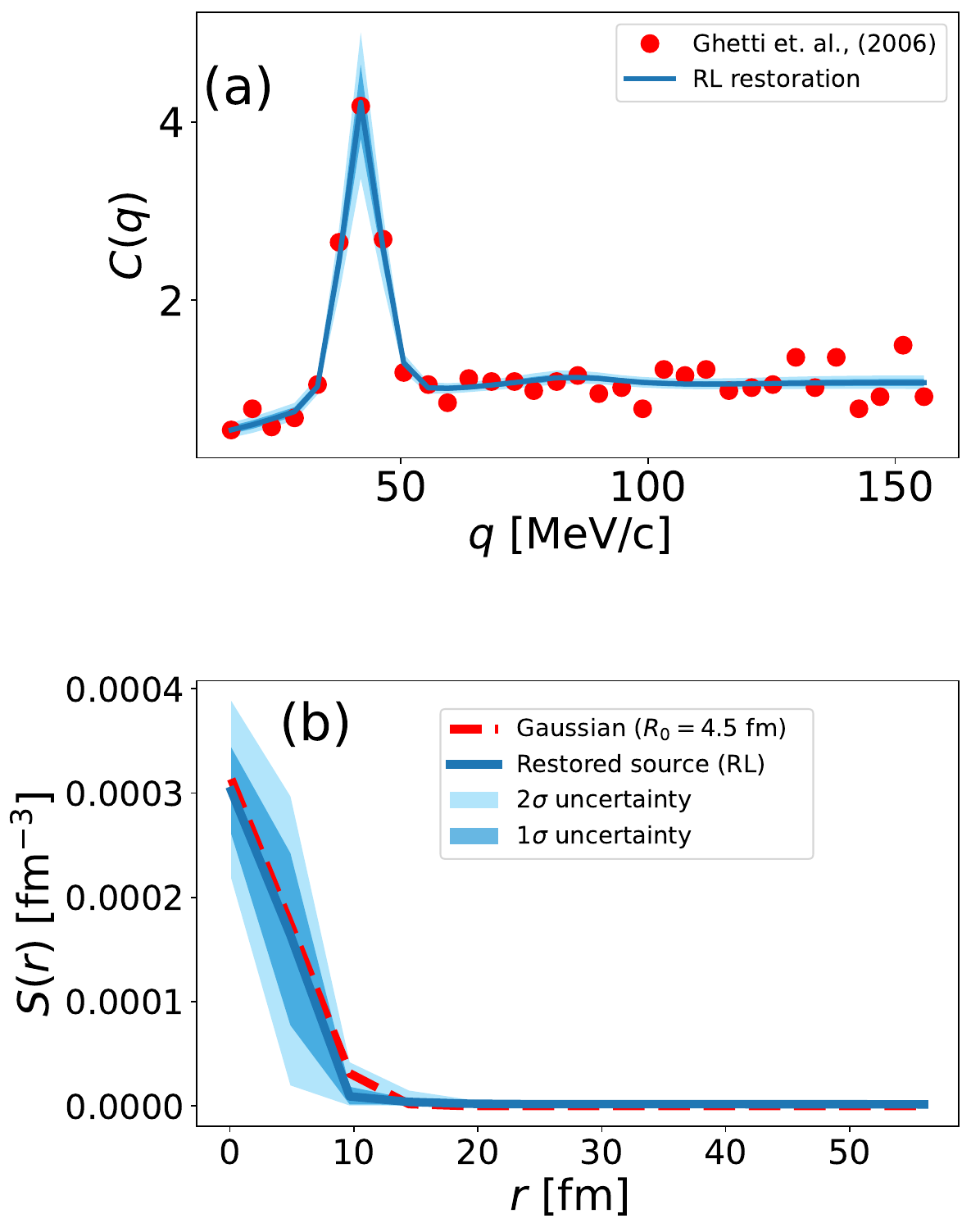}
    \caption{(a) Deuteron-alpha correlation vs magnitude of relative momentum $\mathbf{q} = \mu (\mathbf{v}_1 - \mathbf{v}_2) $ in the center of mass of the pair.  Points represent the correlation measured in the $E/A=44 \, \text{MeV}$ $^{40}$Al+$^{27}$Al reaction by Ghetti {\em et al.}~\cite{GHETTI2006307}.   Line and shaded regions represent results from RL source restoration.  (b) Source inferred from the measured correlation function in (a). The solid line and shaded areas represent results from RL source restoration.  For comparison, a Gaussian source function with radius $R_0 = 4.5 \, \text{fm}$ is shown too. }
    \label{fig:my_label_SF}
\end{figure}

Typical correlation function from measurements \cite{GHETTI2006307} is shown in the panel (a) of Fig.~\ref{fig:my_label_SF}.  The binning in relative momentum and scatter of points is quite typical for the measured light particle correlations in heavy-ion collisions.  The pronounced resonance peak at $q \sim 40 \, \text{MeV}/c$ represents the formation and decay of a $J^\pi=3^+$ ($d$-wave) resonance in $^6$Li and a broad hump around $q \sim 80 \, \text{MeV}/c$ is tied to two higher overlapping resonances with $J^\pi=2^+$ and $J^\pi=1^+$.  Opportunity of observing resonance structures is one of the reason of reaching to correlations as a source of information about interactions in different channels.  The case of $^6$Li is on its own of particular interest as the $^6$Li abundance can inform about the evolution of the Universe \cite{PhysRevLett.113.042501}.  The suppression of the correlation function at low $q$ is due to the $d$--$\alpha$ Coulomb repulsion.
Next, we use the features of the particular measurement as a general guidance in testing the capabilities of the deblurring algorithm in source restoration.  Ahead of the restoration from data, we carry out tests where we first apply a forward relation between assumed source and correlation function.

For the sake of deblurring, the source is discretized.  To fix attention we take the source distance range limited from above by $r_\text{max} = 56 \, \text{fm}$ and divide it into $M$ even bins and represent an isotropic $S$ in the form  
 \begin{eqnarray}
 S(r)=\sum_{j=1}^N S_j \, g_j(r) \, ,
 \label{source}
 \end{eqnarray}
 where $g_j$ is a characteristic function for the $j$'th bin,
 \begin{eqnarray}
g_j(r) = \begin{cases} 1 \, , & \text{if $r_{j-1/2} < r < r_{j+1/2}$} \, ,\\
0 \, , & \text{otherwise.}
\end{cases}
 \end{eqnarray}
 
 We compute the wave functions needed for the relations \eqref{eq:CPsiS},  \eqref{eq:CPsiS1D} and \eqref{eq:KqrPsi}, of the source with the correlation employing the interaction potentials fitted to the $d$-$\alpha$ phase shifts from the measurements by McIntyre and Haeberli \cite{mcintyre_phase_1967}.  Those potentials have been modified \cite{alphascorrelation} relative to those by Boal and Shillcock \cite{PhysRevC.33.549} for a better fit.  With spins made explicit and $u_{\ell J}$ representing radial angular wave function for orbital angular momentum $\ell$ and total $J$, the angle-average wave function is 
 \begin{eqnarray}
    \overline{|\Psi^{(-)}_\mathbf{q} (\mathbf{r})|^2} = \frac{1}{3 (qr)^2} \sum_{\ell \, J} (2J+1) \, |u_{q \ell J}(r)|^2 \, .
    \label{eq:AngleAveraged}
\end{eqnarray}
At modest $q$ we account for nuclear interactions only for low $\ell=0, \, 1, \, 2$.

 With the correlation function determined at momenta $q_i$, $i=1, \ldots, N$, the mapping of the correlation onto the blurring problem amounts to $C(q_i) \equiv C_i \leftrightarrow f_i$, $S(r_j) \equiv S_j \leftrightarrow \mathcal{F}_j$ and
 \begin{eqnarray*}
     4 \pi \int_{r_{j-1/2}}^{r_{j+1/2}} dr \, r^2 \, K(q,r) \leftrightarrow P_{ij} \, .
 \end{eqnarray*}
 We generally use fewer points for the source than in the correlation function, $M \le N$.  For the selection of $q$ in Fig.~\ref{fig:my_label_SF}(a) ($N=34$), our source resolution ends up at $r_\text{max}/M \sim 4 \, \text{fm}$ ($M \sim 14$ for $r_\text{max} = 56 \, \text{fm}$).  Details on that will be provided later.

\begin{figure*}
    \centering
     \includegraphics[scale=0.43]{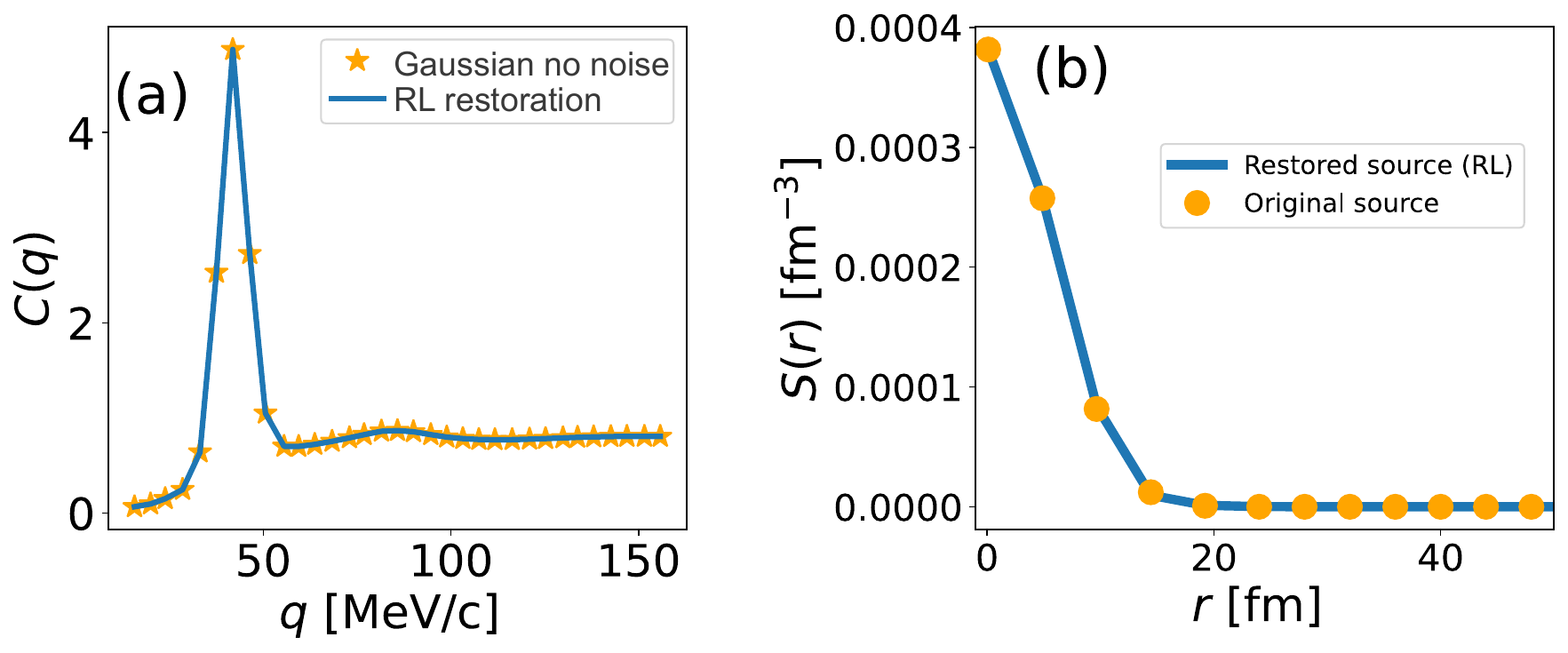}
    \caption{Test of source restoration from a correlation function without noise.  Panel (a) displays the discretized $d$-$\alpha$ correlation function (points) generated from the discretized Gaussian source function displayed in panel (b) (points).  Panel (b) displays further the source restored with RL algorithm (solid line) from the discretized correlation function in (a).  Finally, as a cross-check, panel (a) displays the correlation function (solid line) produced from the restored source.}
    \label{fig:Simulated1}
\end{figure*}

Given the relative success of Gaussian sources in correlation analyses and the features of the particular measured function, we take the source for testing our restoration in the form $S^G(\mathbf{r}) \propto \exp{(-\frac{\mathbf{r}^2}{2R^2_0})}$.  That source with $R_0 = 4.5 \,\text{fm}$ and normalized to 1 is shown with points in the panel~(b) of Fig.~\ref{fig:my_label_SF}.  The correlation function generated with the forward source-correlation relation~\eqref{eq:CPsiS1D} is shown with points in the panel~(a) of Fig.~\ref{fig:Simulated1}.

\begin{figure*}
    \centering
    \includegraphics[scale=0.4]{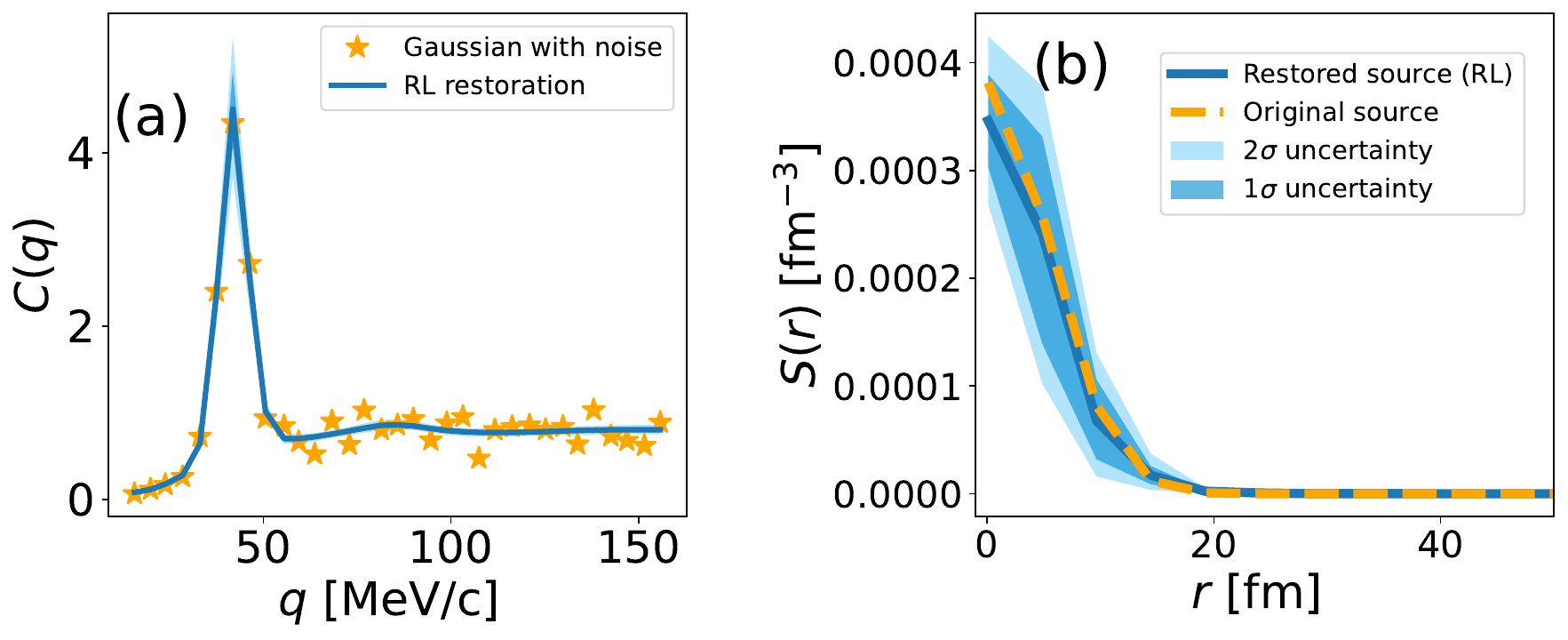}
    \caption{Test of source restoration from a correlation function with noise.  Points in panel (a) represent the correlation function from the model $R=5.5\, \text{fm}$ Gaussian source, with a sample Gaussian noise added.  The Gaussian source itself is represented with a dashed line in panel (b).  The shaded areas and the solid line in panel (b) represent results of restoration from an ensemble of correlation functions such as in (a) where the Gaussian noise was repeatedly sampled.  The solid line shows the average restored source and the darker and lighter shaded areas show the extent of the 1-$\sigma$ and 2-$\sigma$ range in the distribution of restored source function values.  Finally, the solid line and the darker and lighter shaded regions in panel (a) show similar information for the correlation functions from the ensemble of restored source functions.  The narrow width of the 1-$\sigma$ region in (a), compared to the original assumed uncertainties in $C$, stems from the constraint of $S$ being nonnegative, built into the restoration, and $C$ dependent effectively on values of $S$ at just few points in $r$.}
    \label{fig:Simulated2}
\end{figure*}

We test restoration with the RL method both for a smooth and noisy input correlation functions $C(q)$.  The source restored from the smooth function in the panel (a) of Fig.~\ref{fig:Simulated1} is illustrated with lines in the panel (b).  It may be observed that the input and restored source cannot be distinguished within the resolution of the figure.  To test the case of a noisy $C$, see \cite{Pierre}, we add model fluctuations to the smooth $C$. Specifically, we observe that we approximate the scatter of points in the experimental $d$-$\alpha$ correlation function in Fig.~\ref{fig:my_label_SF}(a) around a smooth function $C^\text{sm}$ with a Gaussian characterized by a $q$-dependent width approximately equal to $0.15 \sqrt{C^\text{sm}(q)}$.  With this we sample noisy correlation functions for our tests from $C(q) \sim C^G(q) + 0.15 \sqrt{C^G(q)} \, \mathcal{N}(0,1)$, where $C^G (q)$ is the smooth function generated with the Gaussian source.  A sample correlation function with noise is illustrated with stars in Fig.~\ref{fig:Simulated2}(a).  We generate an ensemble of such correlation functions and the corresponding ensemble of restored sources.  The average values of the restored sources at different $r$ are illustrated with a solid line in Fig.~\ref{fig:Simulated2}(b).  The extent of the 1-$\sigma$ and 2-$\sigma$ ranges in the value distributions for restored sources at different $r$ are illustrated as dark and light shaded areas, respectively.  It can be observed that the restored values generally agree within 1-$\sigma$ with the original Gaussian source.  

It is important to note that the RL algorithm can suffer from noise amplification after a modest number of iterations, see Refs.~\cite{danielewicz2022deblurring, Pierre} and references within. In our calculation, we suppressed potential instability by applying a regularization in the algorithm. The first level of regularization is the binning choice in the source function (see Eq.~\eqref{source}); too many bins lead to oscillations in restoration, and too few bins lead to the loss of information, and we discuss binning choice in detail later in the letter. The second level of regularization, we use here, is the one developed in Ref.~\cite{danielewicz2022deblurring}, where the parameter $\lambda=0.015$ was chosen.

The main goal of our letter remains the restoration of a source from data following the RL algorithm.  The $d$-$\alpha$ pairs yielding the correlation function in Fig.~\ref{fig:my_label_SF} have been measured by Ghetti {\em et al.}~\cite{GHETTI2006307} at forward angles $0.7^\circ<\theta<7^\circ$ in $40 \, \text{MeV/nucl}$ $^{40}$Ar+$^{27}$Al collisions.  When narrow structures are measured in an experiment, such as the $q \sim 40 \, \text{MeV}/c$ peak in the correlation function, then detector resolution needs to be considered.  

The impact of the resolution, as far as the forward relation between the source and correlation is concerned, is in the modification of the kernel in Eq.~\eqref{eq:CPsiS}, where the original kernel $K(\mathbf{q},\mathbf{r})$ gets convoluted with an appropriate detector resolution function pertaining to $\mathbf{q}$.  Relative to the measured vector, the resolution can modify the magnitude of the vector $\mathbf q$ in the wave function, as well as its direction, especially for low $q$.  However, for low values of the $qr$ product, only low $\ell$ will matter in the wavefunction squared in the kernel, so the sensitivity to the $\mathbf q$ direction will be weak.  On the other hand, in the presence of resonances the sensitivity to the pair c.m.\ energy, tied to the detector energy resolution, can be quite strong.  In Ref.~\cite{PhysRevC.36.2297} it has been proposed to account for the smearing in $q$ by folding the original kernel with a Gaussian in $q$ of width $\sigma_q$ adjusted to the energy resolution.  For an angle-averaged correlation function, this yields
\begin{equation}
     K(q,r) = \int d q' \,  \frac{1}{\sqrt{2\pi} \sigma_q} \, \text{e}^{-\frac{(q-q')^2}{2 \sigma_q^2}} \, \overline{|\Psi^{(-)}_{\mathbf{q}'}({\mathbf r})|^2} \, ,
     \label{resolution}
\end{equation}
in Eq.~\eqref{eq:CPsiS1D}.  In the limit of low relative velocity $v$ for the pair, as compared to the velocity of the pair c.m.\ $V$, simple kinematic considerations yield a relation between the resolution in energy $\sigma_E$ and that in relative velocity $\sigma_v$, for angle-averaged $\mathbf q$: $\sigma_v = \frac{1}{\sqrt{6}} \, \frac{\sigma_E}{E} \, V$.  With the energy resolution in the particular experiment $ \frac{\sigma_E}{E} \sim 2\%$ \cite{lanzano_using_1992} and~$V$ corresponding to the projectile-like fragments~\cite{GHETTI2006307}, we get $\sigma_q = \mu \, \sigma_v \simeq 3 \, \text{MeV}/c$.  

\begin{figure*}
    \centering
    \includegraphics[scale=0.5]{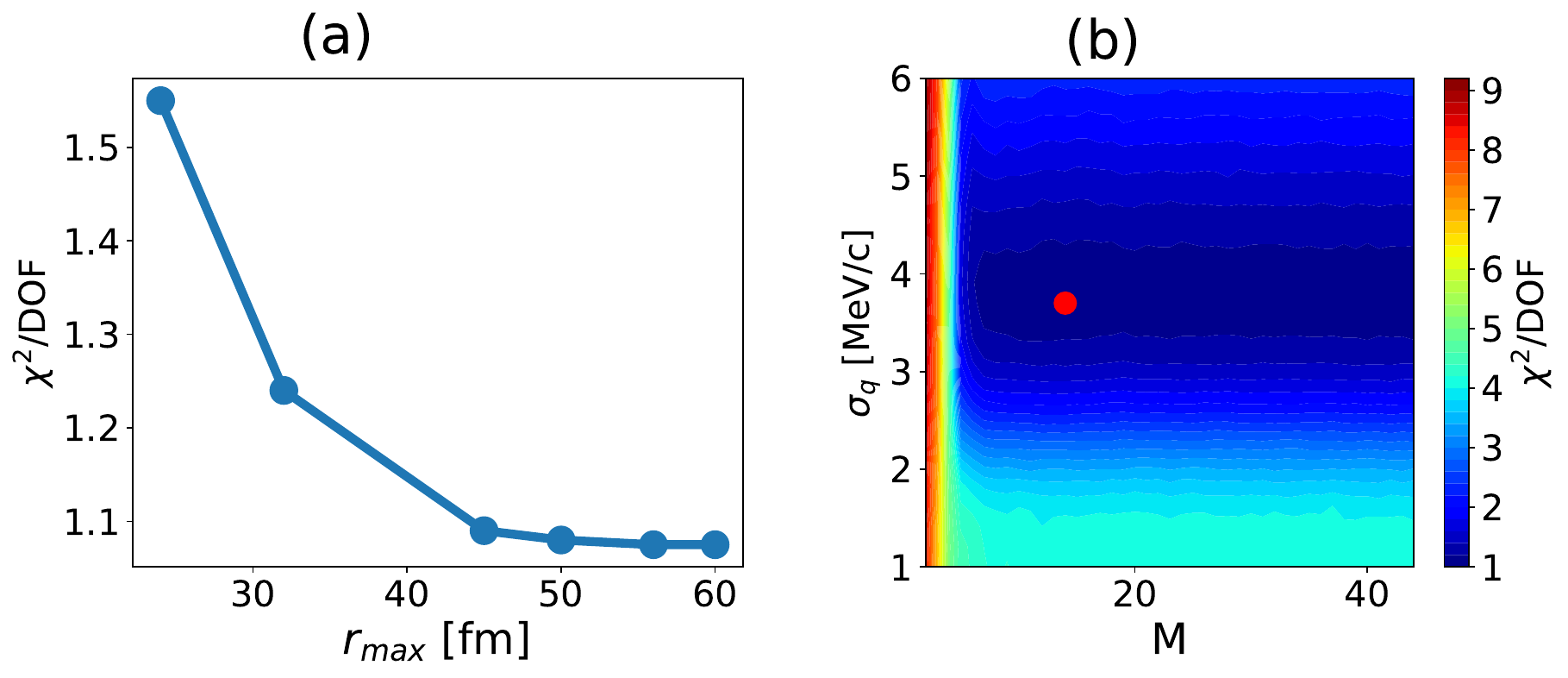}
    \caption{ $\chi^2$ per degree-of-freedom results in describing the measured $d$-$\alpha$ correlation function in terms of the correlation from an RL restored source.  Panel (a) shows minimal $\chi^2/\text{DOF}$ for $\sigma_q$ and $M$ optimized, while $r_\text{max}$ is set.  Panel (b) shows a contour plot of $\chi^2/\text{DOF}$ when $r_\text{max}=56 \, \text{fm}$ and $\sigma_v$ and $M$ are varied.  The circle marks the minimum at $\sigma_q = 3.7 \, \text{MeV}/c$ and $M=14$.
%    
%    Panel (a) displays a contour plot of the chi-squared per degree of freedom ($\chi^2/$DOF), which is used to compute the optimal parameters for the analysis. These parameters include the number of bins (M) in the discretized source and the energy resolution ($\sigma_q$). The chi-squared per degree of freedom is calculated using Eq.\eqref{chi}, for a two-particle source function that spans a relative distance range of 0-56 fm. The red dot on the graph indicates the location of the minimum $\chi^2/$DOF, and the corresponding values of $\sigma_q$ and M are 3.7 MeV/c and 14, respectively. In panel (b), the graph illustrates how $\chi^2/$DOF changes with the range in the relative distance where the source function is defined. 
    }
    %{Contour plot shows the energy integrated square error $E$, a metric used to estimate the best parameters (Resolution and bins). The vertical axis represent the resolution in the correlation function and horizontal axis represents the number of bins of the discretized source function. The red dot shows the position of the best parameters.}
    \label{fig:my_labe2}
\end{figure*}

An alternative to estimating $\sigma_q$ using previously inspected $\sigma_E/E$ is to assess $\sigma_q$ directly from the correlation measurement when a narrow resonant state is present such as for $d$-$\alpha$.  In the source inference from data it is also necessary to decide on the source discretization, i.e., $r_\text{max}$ and $M$ in our scheme. For this, we compare the correlation $C^{RL}$ from the source $S^{RL}$ inferred through RL deblurring to the measured correlation $C$ and construct~$\chi^2$:
\begin{eqnarray}
\chi^2=\sum_{i=1}^N \Big(\frac{ C^{RL}_i-C_i}{\epsilon_i}\Big)^2 \, .
\label{chi}
\end{eqnarray}
Here, the uncertainties are estimated as $\epsilon_i \approx 0.15 \sqrt{C_i}$.  At fixed $r_\text{max}$, $\sigma_q$ and $M$, we carry out the RL deblurring and then minimize $\chi^2$ under variation of $\sigma_q$ and $M$.  With the number of degrees of freedom (DOF) calculated as $N$--$M$--2, where $M$ is for the number of source bins, and~2 is for $M$ and $\sigma_q$ adjustments, we show in Fig.~\ref{fig:my_labe2}(a) $\chi^2/\text{DOF}$ obtained in this manner as a function of $r_\text{max}$.  We find a flat behavior of $\chi^2/\text{DOF}$ at $r_\text{max} \gtrsim 50 \, \text{fm}$, close to 1 for the particular parametrization of $\epsilon$.  In Fig.~\ref{fig:my_labe2}(b) we show a contour plot of $\chi^2/\text{DOF}$ at fixed $r_\text{max} = 56 \, \text{fm}$ when $\sigma_q$ and $M$ are varied.  We typically find the minimum at $\sigma_q \sim 3.8 \, \text{MeV}/c$ and $r_\text{max}/M \sim 4 \, \text{fm}$ at different~$r_\text{max}$.

In the source restoration illustrated in Fig.~\ref{fig:my_label_SF}, we use $r_\text{max}=56 \, \text{fm}$, $M=14$ and $\sigma_q = 3.7 \, \text{MeV}/c$.  The proximity of $\sigma_q$ from the fit to that from the resolution estimate should be noted.  Compared to the Gaussian source there, that restored approaches faster low values by $r \sim 10 \, \text{fm}$, but then it has higher values above $\sim 15 \, \text{fm}$.  Notably, when an unnormalized Gaussian source parameterization is used to describe correlations it gets combined with Eq.~\ref{eq:RKS} or angle-averaged version thereof.  In the latter case, it is assumed that the strength completing source normalization is located at large $r$.

Summing up, we have demonstrated the use of a deblurring method, successful in optics, to infer the emission source from low relative-velocity correlation function.  As the example, we chose the $d$-$\alpha$ correlation that features a narrow and overlapping broad resonances, as well as Coulomb depletion at low $q$.  The source inference involves determination of relative wave function in order to generate the kernel for the KP relation.  In parallel to the binning of the correlation function typical for experiment, the source and, correspondingly, the kernel gets discretized, yielding a transfer matrix.  Impact of detector resolution can be accounted for in the matrix, in parallel to the physics connecting the source and correlation function.  The source restoration progresses through RL iterations until source stabilization. Uncertainties in source determination can be assessed by resampling the experimental correlation function with experimental uncertainties. 

We have tested the source restoration from a $d$-$\alpha$ correlation function with a Gaussian source, both for an idealized function without uncertainties and with uncertainties.  In both cases, an application of the KP relation followed by the RL deblurring returned a source information consistent with the input.

In analyzing the measured $d$-$\alpha$ correlation, we demonstrated that, for sharp resonances, the impact of detector resolution may be read off from the correlation itself.  The source restored from the data through RL deblurring is close, within restoration resolution, to a Gaussian source in the central part, but it first approaches low values more abruptly, to then exhibit a tail that the Gaussian source lacks.

We hope that the RL or other optical deblurring algorithms, applied as here, may turn out being useful in inferring the sources from correlation measurements.

This work was supported by the U.S.\ Department of Energy Office of Science under Grant No.\ DE-SC0019209.
\nocite{}
\bibliographystyle{apsrev}
\bibliography{reference.bib}
\end{document}